\documentclass[twocolumn,showpacs,preprintnumbers,amsmath,amssymb,prb]{revtex4}
\usepackage{amsmath,amsfonts,bm,graphicx}
\usepackage{color}

\begin{document}

\title{Tight-binding study of the magneto-optical properties of gapped graphene}

\author{Jesper Goor Pedersen and Thomas Garm Pedersen}
\affiliation{Department of Physics and Nanotechnology
             Aalborg University, Skjernvej 4A
             DK-9220 Aalborg East, Denmark}

\date{\today}

\begin{abstract}
We study the optical properties of gapped graphene in presence of a magnetic field. We consider a model
based on the Dirac equation, with a gap introduced via a mass term, for which analytical expressions
for the diagonal and Hall optical conductivities can be derived. We discuss the effect of the mass
term on electron-hole symmetry and $\pi$-$\pi^*$ symmetry and its implications for the optical Hall
conductivity. We compare these results with those obtained using a tight-binding model, in which the mass is
modeled via a staggered potential and a magnetic field is included via a Peierls substitution. Considering
antidot lattices as the source of the mass term, we focus on the limit where the mass term dominates the
cyclotron energy. We find that a large gap quenches the effect of the magnetic field. The role of
overlap between neighboring $\pi$ orbitals is investigated, and we find that the overlap has pronounced
consequences for the optical Hall conductivity that are missed in the Dirac model.
\end{abstract}


\maketitle

\section{Introduction}
While already the subject of a Nobel Prize in physics, research in graphene,\cite{Geim2007,Geim2009} 
a single two-dimensional sheet of carbon first
isolated in 2004,\cite{Novoselov2004} seems to show no signs of slowing down. Initial studies have focused on the unique
electronic properties of \emph{pristine} graphene,\cite{Neto2009} such as, e.g., its extreme electron mobility,\cite{Morozov2008}
and a remarkably large cyclotron gap, which has led to integral quantum Hall measurements at room 
temperature.\cite{Novoselov2007} This feature of graphene arises due to its linear band structure near the Fermi
energy, which leads to an unconventional half-integer quantum Hall effect due to the existence of a Landau level
at zero energy.\cite{Novoselov2005,Zhang2005,Gusynin2005,Gusynin2005a}
The field has recently matured to a point where a large part of the focus has shifted to the possible applications of graphene,
in fields as diverse as transistors,\cite{Lin2010} solar cells,\cite{Park2010} hydrogen storage,\cite{Patchkovskii2005} and touchscreen devices.\cite{Bae2010}
This inevitably draws light to one of the serious drawbacks of graphene, namely its semi-metallic
nature. Several ways of opening a gap in graphene have been put forth. If graphene is cut in narrow ribbons, so-called
graphene nanoribbons, quantum confinement effects will induce a band gap, the size of which depends on the width of
the ribbon as well as the intricacies of the ribbon edges.\cite{Nakada1996,Son2006,Brey2006}
Adsorption on graphene of hydrogen has proven to be a very efficient way of inducing substantial band gaps in graphene,
with fully hydrogenated graphene, termed graphane, exhibiting a band gap of several electron volts.\cite{Sofo2007,Elias2009}
We have previously suggested another means of rendering graphene semiconducting, by creating a periodic array of
antidots, termed graphene antidot lattices.\cite{Pedersen2008,Pedersen2008a} The source of the antidots may be actual perforations
of the graphene sheet as envisioned in the original proposal, or it may be, e.g., patterned hydrogen adsorption.\cite{Balog2010}
Transistors based on graphene antidot lattice have successfully been fabricated and characterized.\cite{Bai2010,Kim2010}

Irrespective of the specific mechanism responsible for the band gap, the
simplest and most general way of 
modeling it is through a mass term. Carriers near the Fermi level of graphene can be described via a Hamiltonian that
resembles the Dirac Hamiltonian of massless neutrinos,\cite{Semenoff1984} which has led to the term massless Dirac 
fermions being applied to the low-energy excitations of pristine graphene. Adding a term that acts oppositely on each sublattice 
is equivalent to adding a mass to these otherwise massless quasiparticles and consequently induces a band gap
of twice the magnitude of the mass term. This results in so-called \emph{gapped graphene}.

While the focus of much research is now on inducing a band gap in graphene, the field remains at a point where the fundamental
properties of the gapped structures remain to be investigated. With this in mind, we focus in this paper on
investigating the magneto-optical properties of gapped graphene. Previous studies of such properties have considered 
excitonic effects as the source of the mass term,\cite{Gusynin2006,Gusynin2007} we have in mind structures such as graphene antidot 
lattices, for which substantial band gaps can be induced. Focus will thus largely be on the limit in which the mass term dominates 
or is comparable to the cyclotron energy. While the Dirac equation provides a reasonable approximation of the low-energy structure 
of graphene, we extend previous studies of magneto-optical properties by comparing with results of tight-binding models. 
Of particular interest is the effect of including the overlap between neighboring $\pi$ orbitals, which is of the order $s=0.1$ and
thus hardly insignificant. It is well-known that the overlap breaks electron-hole symmetry in the spectrum of
graphene,\cite{Neto2009} which is of significant consequence for the magneto-optical properties of gapped graphene.
In particular, we demonstrate that the broken symmetry results in a much richer structure of the optical Hall conductivity of
gapped graphene. Also, a non-zero optical Hall conductivity is found, even for a chemical potential sitting in the middle of the gap,
where the electron-hole symmetry inherent in the other models dictates a vanishing optical Hall conductivity.

The article is organized as follows. In section \ref{Models} we first present the Dirac model, and discuss
some intriguing features of the Dirac model related to valley asymmetry before presenting analytical expressions
for the optical conductivities. These expressions were first derived in a slightly different manner
by Gusynin \emph{et al.} in Ref.~\onlinecite{Gusynin2007}. We then present the tight-binding model, emphasizing
the role of overlap between neighboring $\pi$ orbitals. We generalize previous results regarding the role of overlap
in pristine graphene to the case of gapped graphene. In section \ref{Results} we present the results of
the different methods. We first focus on the diagonal optical conductivity and then move on the optical Hall
conductivity. We compare the three different methods and discuss the effect of the mass term, particularly in
the regime where the mass term dominates the cyclotron energy. We illustrate the particular importance of
including overlap in the tight-binding model in relation to the optical Hall conductivity. Finally, 
in \ref{Conclusions} we summarize our findings.

\section{Models}\label{Models}
We consider gapped graphene, i.e. including a mass term, and include
a magnetic field applied perpendicular to the graphene plane, $\mathbf{B}=B\hat{\mathbf{z}}$.
We choose the Landau gauge, $\mathbf{A}=Bx\hat{\mathbf{y}}$, and assume $B>0$ throughout the article.
To calculate the optical response, we use the Kubo-Greenwood formula
\begin{eqnarray}
\frac{\sigma_{\alpha\beta}(\omega)}{\sigma_0} &=& 
-i\frac{4\hbar^2}{Am^2}\nonumber \\
&&\times\sum_{n,m}\frac
{[n(E_{\nu_n}\!)-n(E_{\nu_m}\!)]\Pi^\alpha_{\nu_n\!\nu_m}\Pi^\beta_{\nu_m\!\nu_n}}
{E_{\nu_m,\nu_n}\left[E_{\nu_m,\nu_n}-(\hbar\omega+i\hbar\Gamma)\right]},
\end{eqnarray}
where the sum is over all states, each described by a complete set of quantum numbers, $\nu_n$.
Here, $E_{\nu_m,\nu_n}=E_{\nu_m}-E_{\nu_n}$, while $n(E)$ is the Fermi distribution function, 
$A$ is the unit cell area, $\hbar\Gamma$ represents a broadening term, 
and we have normalized with the zero-frequency graphene conductivity $\sigma_0=e^2/(4\hbar)$. 
Also, we have introduced $\Pi^\alpha_{\nu_n\nu_m}$ as the $\alpha$-component of the canonical momentum matrix
elements 
$\boldsymbol{\Pi}_{\nu_n\!\nu_m}=\left<\nu_n|\boldsymbol{\Pi}|\nu_m\right>$, 
with
$\boldsymbol{\Pi}=\mathbf{p}+e\mathbf{A}$. 
To evaluate the momentum matrix elements we will make use of the
commutator relation 
$\boldsymbol{\Pi}=im \hbar^{-1}[H,\mathbf{r}]$.

In order to make the comparison with tight-binding results more clear, and to illustrate in detail some of the
intriguing properties of the model, we will proceed by deriving analytical results of the optical conductivity of gapped 
graphene, in a model based on the Dirac equation. This will serve to highlight the differences and similarities between
the Dirac and the tight-binding treatment of the problem, and provide us with some analytical expressions with which to
compare the tight-binding results. We stress that the analytical expressions for the optical conductivity
have already been derived previously by Gusynin \emph{et al.} in Ref.~\onlinecite{Gusynin2007}, although with a slightly different
approach than the one we will adopt. We thus repeat some of these results as well as others by
Jiang \emph{et al.}\cite{Jiang2010}, in order to assist in the comparison with the tight-binding results later. 
Also, the derivation will serve to clarify the effect of the mass term on the symmetry between electron and hole states,
which has significant implications for the optical Hall conductivity.

\subsection{Dirac equation}
The low-energy excitations of graphene near 
the $K$ valley are well described by a Dirac Hamiltonian\cite{Semenoff1984} 
$H_K=v_F \boldsymbol{\Pi}\cdot\boldsymbol{\sigma}+\Delta\sigma_z$, with $\boldsymbol{\sigma}=(\sigma_x,\sigma_y)$ where
$\sigma_i$ are Pauli spin matrices. Here, we have included a mass term $\Delta \sigma_z$, which breaks the sublattice symmetry
and adds a band gap to graphene of size $2\Delta$.
The two inequivalent valleys of graphene are related via time-reversal
symmetry and a Hamiltonian valid near the $K^\prime$ valley is thus obtained
by substituting $\boldsymbol{\sigma}\rightarrow \boldsymbol{\sigma}^*$. 
In matrix notation, the Hamiltonian thus reads
\begin{equation}
H=\left[	
\begin{array}{cc}
\Delta & \Pi_\mp \\
\Pi_\pm & -\Delta
\end{array}
\right].
\end{equation}
Here, the upper (lower) sign corresponds to the $K$ ($K^\prime$) valley, and we have introduced
$\Pi_\pm = v_F(p_x \pm ip_y \pm ieBx)$, with $v_F$ the Fermi velocity of graphene. 
The eigenvalues of this Hamiltonian are\cite{Jiang2010}
\begin{eqnarray}
E_n^K &=& \mathrm{sgn}(n)\sqrt{\Delta^2+\hbar^2\omega_c^2|n|}-\delta_{n0}\Delta,\;\;n=0,\pm1,\ldots \nonumber \\
E_n^{K^\prime} &=& \mathrm{sgn}(n)\sqrt{\Delta^2+\hbar^2\omega_c^2|n|}+\delta_{n0}\Delta,\;\;n=0,\pm1,\ldots \nonumber \\
\end{eqnarray}
where $\omega_c=\sqrt{2}v_F/l_B$, with the magnetic length $l_B=\sqrt{\hbar/(eB)}$, and we define $\mathrm{sgn}(0)=0$.
This results suggest that in the presence of a mass term, particle-hole symmetry is no longer retained in
each individual valley, where it is broken by the single $E_0=\pm\Delta$ eigenstate.\cite{Jiang2010} 
In addition to the absence of
particle-hole symmetry in the energy spectrum, we find that the general $\pi\!-\!\pi^*$ symmetry between electron-hole pairs of
eigenstates is broken, because the mass term breaks sublattice symmetry. 
Letting $A_{nk}^\kappa$ and $B_{nk}^\kappa$ denote the individual spinor components of the eigenstates, we find that
while in the absence of a mass term,
the $\pi\!-\!\pi^*$ symmetry relates a given hole state to its corresponding electron state via
$\Psi_{-n,k}^K=(-A_{nk}^K, B_{nk}^K)^T$, the mass term results in a relation instead between states in opposite valleys,
$\Psi_{-n,k}^K=(B_{nk}^{K^\prime}, A_{nk}^{K^\prime})^T$. We will see that this has particular consequences for
the optical Hall conductivity in gapped graphene.

To evaluate the momentum matrix elements we note that the commutator relation yields
$\boldsymbol{\Pi}^K=mv_F\boldsymbol{\sigma}$ while $\boldsymbol{\Pi}^{K^\prime}=(\boldsymbol{\Pi}^K)^*$.
This yields the transition rule $|m|=|n|\pm 1$ for optical transitions $\left|n\right>\rightarrow\left|m\right>$.
\begin{figure}
\begin{center}
\includegraphics[width=.7\linewidth]{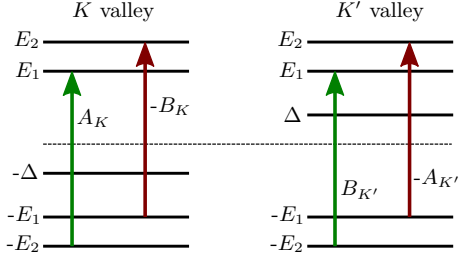}
\caption{(Color online) Transitions contributing to the optical Hall conductivity in gapped graphene,
in the case of zero temperature and chemical potential. The weights of each transition is indicated,
with color indicating the sign. Note that for vanishing gap, $A_\kappa=B_\kappa$ and the Hall conductivity
vanishes within each valley. This does not hold in the gapped case, where instead contributions from opposite valleys
cancel.
}
\label{fig:DEtrans}
\end{center}
\end{figure}
We find that the optical conductivity tensor can then be written in the form
\begin{eqnarray}
\frac{\sigma_{x\beta}(\Omega)}{\sigma_0} &=& 
(-i)^{\delta_{\beta\!,x}}3\phi
\nonumber\\
&\times&
\sum_{n,m}
\left[
\delta_{|n|-1,|m|}
\left(A_{K}(n,m)+B_{K^\prime}(n,m)\right)
\right.
\nonumber \\
&+&\left.\!\!\!\!
(-1)^{\delta_{\beta\!,y}} \delta_{|n|,|m|-1} 
\left(B_{K}(n,m)\!+\!A_{K^\prime}(n,m)\right)
\right]
,\label{eq:sigmaDE}
\end{eqnarray}
where all energies are now in units of the hopping element $t$, 
$\Omega=\hbar\omega/|t|$ and $\gamma=\hbar\Gamma/|t|$,
and we have introduced the relative 
magnetic flux $\phi=Ba^2/\Phi_0$, with the magnetic flux quantum $\Phi_0=\pi\hbar/e$. Also, we define
\begin{eqnarray}
A_\kappa(n,m) &=& 
\frac
{\left[n(\epsilon_n^\kappa)-n(\epsilon_m^\kappa)\right]|a_n^\kappa|^2|b_m^\kappa|^2}
{(\epsilon_m^\kappa-\epsilon_n^\kappa)\left[(\epsilon_m^\kappa-\epsilon_n^\kappa)-(\Omega+i\gamma)\right]},\nonumber \\
B_\kappa(n,m) &=& 
\frac
{\left[n(\epsilon_n^\kappa)-n(\epsilon_m^\kappa)\right]|b_n^\kappa|^2|a_m^\kappa|^2}
{(\epsilon_m^\kappa-\epsilon_n^\kappa)\left[(\epsilon_m^\kappa-\epsilon_n^\kappa)-(\Omega+i\gamma)\right]},\label{eq:ABkappa}
\end{eqnarray}
where $\epsilon_n^\kappa=E_n^\kappa/|t|$, with $\kappa\in\{K,K^\prime\}$. Also, we have introduced
$a_n^\kappa=\mathrm{sgn}(E_n^\kappa+\Delta)\sqrt{(1+\Delta/E_n^\kappa)/2}$ and
$b_n^\kappa=i\sqrt{(1-\Delta/E_n^\kappa)/2}$. 
In terms of the relative
magnetic flux $\epsilon_n^\kappa=\mathrm{sgn}(n)\sqrt{\delta^2+(3\pi/2)\phi|n|}\mp \delta_{n0}\delta$.
Note that the expressions in Eq.~(\ref{eq:ABkappa}) only differ for the different valleys 
if the $n=0$ state is involved in the transition.

The expression Eq.~(\ref{eq:sigmaDE}) for the optical conductivity highlights an interesting feature of gapped graphene.
Ordinarily, at zero temperature and chemical potential, $\pi\!-\!\pi^*$ symmetry means that contributions from transitions 
$\left|-n\right>\!\rightarrow\!\left|m\right>$ and $\left|-m\right>\!\rightarrow\!\left|n\right>$ 
exactly cancel in the sum for the off-diagonal optical conductivity.\cite{Pedersen2003} Letting $\Delta=0$ we indeed find
$|a_n|^2=|b_n|^2$ and thus an exact cancellation of terms in the sum in Eq.~(\ref{eq:sigmaDE}), even if restricted
to just a single valley.
However, for $\Delta\neq 0$ this equality no longer holds and the lack of $\pi\!-\!\pi^*$ symmetry 
results in a non-zero optical Hall conductivity within each individual valley, as discussed also in the DC
case by Jiang \emph{et al}.\cite{Jiang2010}
Instead, the modified symmetry induced by the mass term between electron and
hole states in \emph{opposite} valleys means that contributions
from transitions
$\left|-n\right>\!\rightarrow\!\left|m\right>$ in one valley
are canceled by transitions $\left|-m\right>\!\rightarrow\!\left|n\right>$ 
in the opposite valley,
resulting in the expected $\sigma_{xy}(\omega)=0$ at zero temperature and chemical potential, as illustrated
in Fig.~\ref{fig:DEtrans}. It it thus crucial to take into account both valleys, and to treat
the asymmetry of the valleys with respect to energy spectrum and eigenstates properly.

In the general case, summing contributions from both valleys,
the diagonal optical conductivity can be written
\begin{widetext}
\begin{eqnarray}
\frac{\sigma_{xx}(\omega)}{\sigma_0} &=&
-i3(\Omega+i\gamma)\phi 
\times
\sum_{n=0}^\infty 
\left[
\left(1-\frac{\delta^2}{\epsilon_n\epsilon_{n+1}}\right)\times
\frac{\left[n(-\epsilon_{n+1})-n(-\epsilon_n)\right]+\left[n(\epsilon_n)-n(\epsilon_{n+1})\right]}
{\left(\epsilon_{n+1}-\epsilon_n\right)\left[\left(\epsilon_{n+1}-\epsilon_n\right)^2-\left(\Omega+i\gamma\right)^2\right]}
\right. \nonumber \\
&& \left.
+
\left(1+\frac{\delta^2}{\epsilon_n\epsilon_{n+1}}\right)\times
\frac{\left[n(-\epsilon_{n+1})-n(\epsilon_n)\right]+\left[n(-\epsilon_n)-n(\epsilon_{n+1})\right]}
{\left(\epsilon_{n+1}+\epsilon_n\right)\left[\left(\epsilon_{n+1}+\epsilon_n\right)^2-\left(\Omega+i\gamma\right)^2\right]}
\right],
\end{eqnarray}
while the optical Hall conductivity reads
\begin{eqnarray}
\frac{\sigma_{xy}(\omega)}{\sigma_0} &=& 3\phi\sum_{n=0}^\infty 
\left(
\left[n(-\epsilon_{n+1})-n(-\epsilon_n)\right]-\left[n(\epsilon_{n})-n(\epsilon_{n+1})\right]\right)
\nonumber \\
&&
\times\left[
\left(1-\frac{\delta^2}{\epsilon_n \epsilon_{n+1}}\right)
\frac{1}{(\epsilon_{n+1}-\epsilon_n)^2-(\Omega+i\gamma)^2}
+
\left(1+\frac{\delta^2}{\epsilon_n \epsilon_{n+1}}\right)
\frac{1}{(\epsilon_{n+1}+\epsilon_n)^2-(\Omega+i\gamma)^2}
\right],
\end{eqnarray}
\end{widetext}
where $\delta=\Delta/t$ and we have defined $\epsilon_0=\delta$. These results are in agreement with those of 
Gusynin \emph{et al.} in Ref.~\onlinecite{Gusynin2007}, while derived in a different manner. We state them again
here for completeness.

\subsection{Tight-binding}
\begin{figure}
\begin{center}
\includegraphics[width=\linewidth]{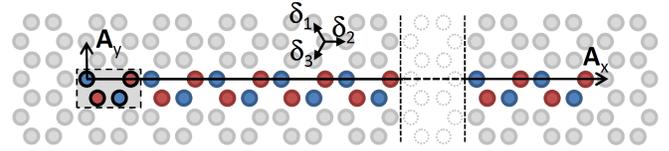}
\caption{(Color online) Unit cell used in the tight-binding calculations. The colored circles indicate the carbon atoms
included in the unit cell, with red and blue indicating different sublattices. The dashed rectangle indicates the
fundamental unit cell, which is repeated in order to ensure translational symmetry. The lattice vectors of the enlarged
unit cell are denoted $\mathbf{A}_x=L_x \hat{\mathbf{x}}$ and $\mathbf{A}_y=L_y \hat{\mathbf{y}}$. Also shown are
the three nearest neighbor vectors $\boldsymbol{\delta}_n$.
}
\label{fig:unitCell}
\end{center}
\end{figure}

Graphene is relatively well described in a nearest-neighbor tight-binding (TB) framework, considering only
$\pi$-electrons. While commonly $\pi$-orbitals $\varphi(\mathbf{r-\mathbf{R}})$ centered on different
sites $\mathbf{R}$ are assumed orthogonal, we will consider also
the case where the overlap between neighboring orbitals is included via the overlap integral
$s=0.1$. This turns out to have a significant impact on the off-diagonal term of the optical conductivity
tensor. 
To include the effects of a magnetic field, we use the Peierls substitution, i.e.
we adopt a minimum coupling substitution 
$\mathbf{p}\rightarrow\mathbf{p}+e\mathbf{A}$ and choose the basis
set $\left<\mathbf{r}|\mathbf{R}\right>=e^{i\phi(\mathbf{R},\mathbf{r})}\varphi(\mathbf{r}-\mathbf{R})$,
where the phase factor is given as a line integral\cite{Luttinger1951}
\begin{equation}
\phi(\mathbf{R},\mathbf{r})=\frac{e}{\hbar}\int_\mathbf{R}^\mathbf{r} \mathbf{A}\cdot d\mathbf{l}.
\end{equation}
In this basis, the Hamiltonian matrix elements can then be approximated as
\begin{equation}
\left<\mathbf{R}|H|\mathbf{R^\prime}\right>=\left<\mathbf{R}|H_0|\mathbf{R^\prime}\right>e^{i\phi(\mathbf{R},\mathbf{R^\prime})},
\end{equation}
where $H_0$ denotes the Hamiltonian in absence of a magnetic field.
In a coordinate system where the $x$ axis is aligned with carbon bonds, see Fig.~\ref{fig:unitCell}, the phase factor becomes
\begin{equation}
\phi(\mathbf{R},\mathbf{R^\prime}) = \frac{\pi}{2}\frac{B}{\Phi_0}\left(x+x^\prime\right)\left(y^\prime-y\right).
\end{equation}
Because of the $(x+x^\prime)$ term, we are forced to choose a unit cell substantially larger than the Wigner--Seitz cell
of graphene in order to retain periodicity in the problem. In particular, we use a rectangular unit
cell of area $L_x \times L_y$, with $L_x=4\Phi_0/(Ba)$ and $L_y=a$, where $a=2.46$~\AA~is the graphene lattice constant.
The unit cell is illustrated in Fig.~\ref{fig:unitCell}.
Letting $N$ denote the number of atoms in the magnetic unit cell, we have 
$N=16\Phi_0/(\sqrt{3}a^2)\times B^{-1}\simeq 316~10^3~\mathrm{T}\times B^{-1}$, illustrating the disadvantage of
the tight-binding approach, namely that very large unit cells are required in order to simulate realistically small
magnetic field strengths. 

Analogously to the case of the Dirac equation, a gap can be introduced in the tight-binding description 
by adding a staggered potential with opposite sign on each sublattice, i.e. taking diagonal elements
$\left<\mathbf{R}|H|\mathbf{R}\right>=\epsilon_\pi+(-1)^n\Delta$, with $n=0,1$ for orbitals sitting on the A and
B sublattice, respectively. Here, for generality, we have included the on-site energy $\epsilon_\pi$ of 
the $\pi$-orbitals.

If we take into account the overlap between neighboring $\pi$-orbitals, the problem becomes that of a
generalized eigenvalue problem. However, advantage can be taken of the similar form of the Hamiltonian
$\hat{\mathbf{H}}$ and the overlap matrix $\hat{\mathbf{S}}$ in the chosen basis if we consider just
nearest-neighbor coupling terms. In particular, the generalized eigenvalue problem can be written in the form
\begin{equation}
\left[
\!
\begin{array}{cc}
(\epsilon_\pi\!+\!\Delta)\hat{\mathbf{I}} & t\hat{\mathbf{F}}\\
t\hat{\mathbf{F}}^\dagger & (\epsilon_\pi\!-\!\Delta)\hat{\mathbf{I}}
\end{array}
\right]
\left[\begin{array}{c}
\mathbf{v}_{\!A} \\
\mathbf{v}_{\!B} 
\end{array}\right] 
=
E
\left[\begin{array}{cc}
\hat{\mathbf{I}} & s\hat{\mathbf{F}}\\
s\hat{\mathbf{F}}^\dagger &\hat{\mathbf{I}}
\end{array}\right]
\left[\begin{array}{c}
\mathbf{v}_{\!A} \\
\mathbf{v}_{\!B} 
\end{array}\right],
\end{equation}
where $\hat{\mathbf{I}}$ is the identity matrix, the eigenvectors have been separated into components on different sublattices, 
and we have introduced the nearest-neighbor transfer integral $t$, evaluated in the absence of a magnetic field.
The similar form of the matrices allows us to rewrite the problem as an ordinary eigenvalue problem for either
sublattice, and in this way arrive at an equation
\begin{equation}
\frac{\left(E-\epsilon_\pi\right)^2-\Delta^2}{(Es-t)^2} = \frac{\left(E_0-\epsilon_\pi\right)^2-\Delta^2}{t^2},
\end{equation}
relating the eigenvalues $E$ to the eigenvalues $E_0$, obtained by ignoring the overlap. This is readily
solved to yield the explicit relation
\begin{eqnarray}
E &=& \frac{1}{t-s^2\left(\xi_0^2-\Delta^2\right)/t}\times
\left\{
t\epsilon_0-s\left(\xi_0^2-\Delta^2\right)
\right. \nonumber \\
&&\left.
\pm\left[
\left(t\epsilon_0-s\left(\xi_0^2-\Delta^2\right)\right)^2
\right.\right. \nonumber \\
&&\left.\left.
+\left(\xi_0^2-\epsilon_\pi^2\right)\left(t^2-s^2\left(\xi_0^2-\Delta^2\right)\right)
\right]^{1/2}
\right\}\label{eq:EsEs0},
\end{eqnarray}
where $\xi_0=E_0-\epsilon_\pi$. Here, the sign of the square root should follow the sign of the energy $E_0$.
We can thus immediately determine the eigenvalues with the overlap included, without solving an additional generalized
eigenvalue problem. Furthermore, we use the same eigenvectors, only properly orthonormalized according
to the overlap. While using Eq.~(\ref{eq:EsEs0})
to calculate the eigenvalues with overlap is exact, this way of determining the eigenvectors constitutes an approximation.
However, we have confirmed numerically that this has no discernible effect of the calculated optical conductivities so long 
as $s\Delta\ll t$.

In addition to offering a decrease in computation time, the similarities of the $\hat{\mathbf{H}}$ and $\hat{\mathbf{S}}$ 
matrices allows us to greatly simplify the calculation of the momentum matrix elements. Following ideas similar to 
those used by Sandu in Ref.~\onlinecite{Sandu2005} for evaluating the momentum matrix elements of ordinary graphene, we consider
the Hamiltonian of the ordinary eigenvalue problem
$\hat{\mathbf{H}}_S=\hat{\mathbf{S}}^{-1}\hat{\mathbf{H}}$. The inverse of the overlap matrix can be written
\begin{equation}
\hat{\mathbf{S}}^{-1} = 
\left[\begin{array}{cc}
\left(\hat{\mathbf{I}}-s^2\hat{\mathbf{F}}\hat{\mathbf{F}}^\dagger\right)^{-1} 
& 
-s\hat{\mathbf{F}}\left(\hat{\mathbf{I}}-s^2\hat{\mathbf{F}}^\dagger\hat{\mathbf{F}}\right)^{-1}
\\
-s\hat{\mathbf{F}}^\dagger\left(\hat{\mathbf{I}}-s^2\hat{\mathbf{F}}\hat{\mathbf{F}}^\dagger\right)^{-1}
& 
\left(\hat{\mathbf{I}}-s^2\hat{\mathbf{F}}^\dagger\hat{\mathbf{F}}\right)^{-1}
\end{array}\right].
\end{equation}
We now disregard terms with $s^2\ll 1$, so that
\begin{equation}
\hat{\mathbf{H}}_S = 
\left[\begin{array}{cc}
\left(\epsilon_\pi+\Delta\right)\hat{\mathbf{I}}-s t\hat{\mathbf{F}}\hat{\mathbf{F}}^\dagger
&
\left(t-s\left(\epsilon_\pi-\Delta\right)\right)\hat{\mathbf{F}}
\\
\left(t-s\left(\epsilon_\pi+\Delta\right)\right)\hat{\mathbf{F}}^\dagger
&
\left(\epsilon_\pi-\Delta\right)\hat{\mathbf{I}}-s t\hat{\mathbf{F}}^\dagger\hat{\mathbf{F}}
\end{array}\right].
\end{equation}
The form of this matrix makes it evident how the inclusion of overlap effectively results in
next-nearest neighbor coupling.
In the crystal momentum representation, the momentum operator is proportional to 
$\nabla_\mathbf{k} H$, which for the Hamiltonian above becomes
\begin{equation}
\nabla_\mathbf{k}\hat{\mathbf{H}}_S = 
\left[
\begin{array}{cc}
-st\nabla_\mathbf{k}\left(\hat{\mathbf{F}}\hat{\mathbf{F}}^\dagger\right)
&
\left(t-s\left(\epsilon_\pi-\Delta\right)\right)\nabla_\mathbf{k}\hat{\mathbf{F}}
\\
\left(t-s\left(\epsilon_\pi+\Delta\right)\right)\nabla_\mathbf{k}\hat{\mathbf{F}}^\dagger
&
-st\nabla_\mathbf{k}\left(\hat{\mathbf{F}}^\dagger\hat{\mathbf{F}}\right)
\end{array}\right].
\end{equation}
If we now assume $s\Delta \ll t$ and let $\epsilon_\pi=0$ this matrix simplifies to
\begin{equation}
\nabla_\mathbf{k}\hat{\mathbf{H}}_S = 
t\!\!
\left[
\begin{array}{cc}
0
&
\nabla_\mathbf{k}\hat{\mathbf{F}}
\\
\nabla_\mathbf{k}\hat{\mathbf{F}}^\dagger
&
0
\end{array}\right]
-
st\!\!
\left[
\begin{array}{cc}
\nabla_\mathbf{k}\left(\hat{\mathbf{F}}\hat{\mathbf{F}}^\dagger\right)
&
0
\\
0
&
\nabla_\mathbf{k}\left(\hat{\mathbf{F}}^\dagger\hat{\mathbf{F}}\right)
\end{array}\right].
\end{equation}
For gapped graphene in the absence of a magnetic field, the second term is 
proportional to the unit matrix and thus does not contribute to the momentum
matrix elements, meaning that the momentum operator is unaltered by the inclusion
of overlap in the model, as is the case for ordinary graphene with no mass term.~\cite{Sandu2005}
However, in the present case, where a magnetic field is included, such a simplification 
can no longer be made. Instead, we note that the second term relies on two small effects,
that of the overlap and that of the magnetic field. Consequently, we ignore the effect of the 
magnetic field on the matrix elements of $\hat{\mathbf{F}}$ in this term, which allows us to disregard 
the second term entirely, as it no longer contributes to the momentum matrix elements.
Effectively, we are thus ignoring non-orthogonality of our basis set and calculating the
momentum matrix elements via $\boldsymbol{\Pi}_{nm}=i(m/\hbar)H_{nm}(\mathbf{R}_m-\mathbf{R}_n)$.

In the numerical implementation of the tight-binding model we employ a simple equidistant $k$-point mesh.
Our choice of unit cell results in negligible dispersion along $k_x$, so we only discretize along 
the $k_y$ direction, ensuring that the folded high-symmetry points are included in the discretization. 
We have verified that including more than a single $k_x$ point has no influence on the results for any realistic 
values of the magnetic field.

\section{Results and discussion}\label{Results}
We now turn to the results for the optical conductivity. Unless otherwise stated, from hereon we will use
parameters $\epsilon_\pi=0$, $s=0.1$ and $t=-2.2$~eV, we include a broadening term $\hbar\Gamma=0.05$~eV and
fix the temperature at $300$~K. Our choice of the value of the hopping term $t$ is motivated by
the transition energy of the saddle point resonance in ordinary graphene, which occurs at 
$2|t|=4.4$~eV.\cite{Taft1964,Pedersen2003} We note that this value of the hopping term results in a Fermi velocity
that does not match experimental results. Without introducing interaction terms in the form of excitonic
effects, one cannot account both for the location of the saddle point resonance and the proper value
of the Fermi velocity. In the end, the exact choice of hopping term does not have a qualitative influence on 
the results we arrive at.

Because of the inherent problems of the Peierls substitution for the tight binding method, namely the
inconvenient scaling of the unit cell with magnetic field, we are limited to quite large magnetic fields.
Therefore, most results will be for a substantial magnetic field of $78$~T. However, we expect the qualitative
features and the conclusions to be valid at more realistic magnetic fields. This claim will be 
substantiated when we compare the tight binding results to those obtained using the Dirac equation.

\subsection{Diagonal conductivity}
\begin{figure}
\begin{center}
\includegraphics[width=\linewidth]{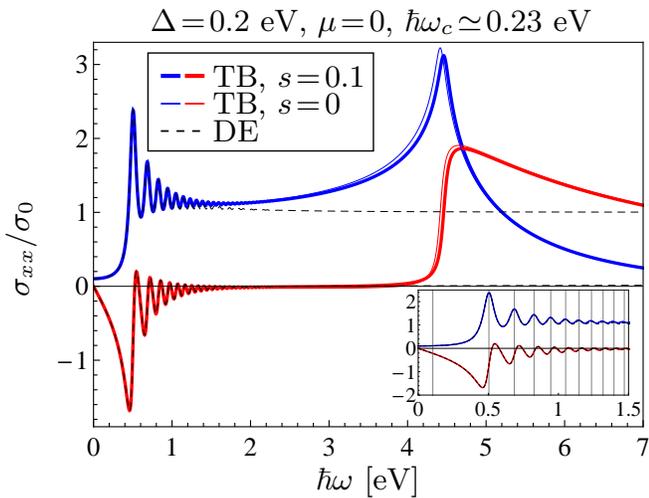}
\caption{(Color online) Diagonal conductivity $\sigma_{xx}$ in units of the DC graphene
conductivity $\sigma_0$, as a function of photon energy $\hbar\omega$. Results are shown for a magnetic field $B=78$~T. Thick lines
indicate results from the TB model including overlap, while thin lines are without overlap.
Dashed lines are DE results. Blue (red) coloring indicates the real (imaginary) part of the conductivity.
Note the resonance around $\hbar\omega=4.4$~eV, which arrises due to the van Hove signularity
at the $M$ point of graphene, and is thus absent in the DE results. The inset offers a closer view of the low-energy oscillations,
illustrating the excellent agreement between all three methods in this energy range. The vertical lines mark the transition energies
predicted by the DE model.
}
\label{fig:xxd0p2mu0}
\end{center}
\end{figure}

We now consider gapped graphene with a mass term $\Delta=0.2$~eV. We fix the chemical potential at midgap, $\mu=0$, and use $T=300$~K. 
In Fig.~\ref{fig:xxd0p2mu0} we show the resulting diagonal optical conductivity calculated using DE and TB with and without overlap. 
The results from all three methods indicate a very clear absorption edge, in excellent agreement with the DE solution, 
which predicts the first resonance peak at
$\hbar\omega_0=\Delta\left(1+\sqrt{1+\hbar^2\omega_c^2/\Delta^2}\right)$. 
Focusing attention on photon energies below $1.5$~eV, we find almost perfect agreement between all three methods, as illustrated in
the inset of Fig.~\ref{fig:xxd0p2mu0}. Here, the resonances fall perfectly on the transition energies predicted by the DE model,
as indicated by the vertical lines in the inset.

However, a notable feature of the full range of energies is the peak near $2|t|=4.4$~eV, which stems
from the van Hove singularity at the $M$ point of graphene. Naturally, such a band structure feature is not reproduced in a DE model.
This $M$-point resonance is unaffected by the addition of a mass term, so long as $\Delta\ll t$. We note that the overlap results in
a slight blueshift of the resonance peak, so that it occurs at $2|t|/(1-s^2)$.

\begin{figure}
\begin{center}
\includegraphics[width=\linewidth]{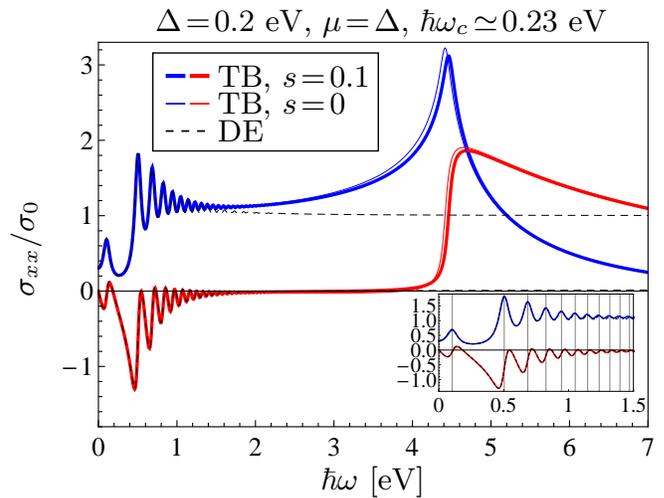}
\caption{(Color online) Diagonal conductivity $\sigma_{xx}$ in units of the DC graphene
conductivity $\sigma_0$, as a function of photon energy $\hbar\omega$. Results are shown for a magnetic field $B=78$~T. 
The chemical potential is set at the value of the mass term, $\mu=\Delta$.
Thick lines
indicate results from the TB model including overlap, while thin lines are without overlap.
Dashed lines are DE results. Blue (red) coloring indicates the real (imaginary) part of the conductivity.
The inset offers a closer view of the low-energy oscillations,
illustrating the excellent agreement between all three methods in this energy range.
The vertical lines mark the transition energies
predicted by the DE model.
}
\label{fig:xxd0p2mu0p2}
\end{center}
\end{figure}
In Fig.~\ref{fig:xxd0p2mu0p2} we show the diagonal optical conductivity in the case where the chemical potential sits on top of 
the lowest Landau level at $\mu=\Delta$. The DE model suggest that in this case the lowest resonance, $\hbar\omega_0$, will
split into two resonances at 
$\hbar\omega_{\pm}=\Delta\left(\sqrt{1+\hbar^2\omega_c^2/\Delta^2}\pm 1\right)$ \cite{Gusynin2007}. Other than this splitting
of the lowest resonance, the results are nearly identical to those obtained with the chemical potential at midgap.
The inset in Fig.~\ref{fig:xxd0p2mu0p2} again shows excellent agreement
between all three methods.

\begin{figure}
\begin{center}
\includegraphics[width=\linewidth]{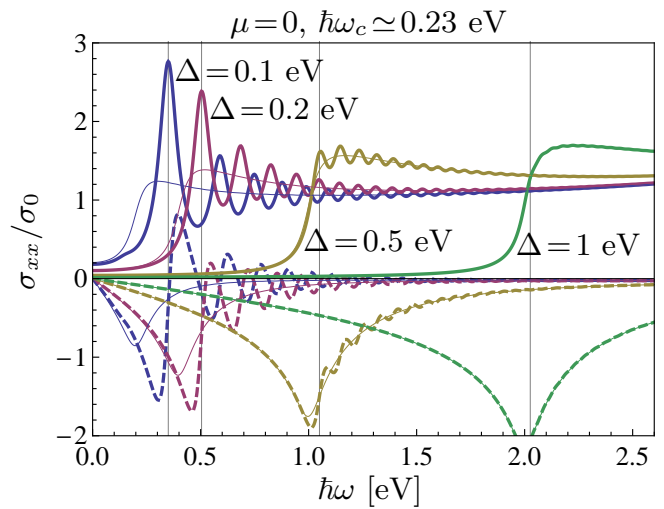}
\caption{(Color online) Diagonal conductivity $\sigma_{xx}$ in units of the DC graphene
conductivity $\sigma_0$, as a function of photon energy $\hbar\omega$. Results are shown for a magnetic field $B=78$~T
and for four different values of the mass term, all calculated using the TB model with overlap. Full (dashed) lines
indicate the real (imaginary) part of the conductivity. For comparison, the thin lines show the optical conductivity 
in absence of a magnetic field. The vertical lines indicate the resonance energies predicted by the DE model.
}
\label{fig:xxmass}
\end{center}
\end{figure}
In Fig.~\ref{fig:xxmass} we illustrate the effect of the mass term on the diagonal conductivity. We show results from the TB model 
with overlap, and with a value of the mass term increasing from $0.1$~eV to $1$~eV. We fix the chemical potential at midgap.
Considering the eigenenergies obtained using the Dirac equation,
$E_n=\Delta(\sqrt{1+(\hbar^2\omega_c^2/\Delta^2) |n|} \pm 1)$,
we expect that the effect of the mass term will be to compress the Landau level spacing, resulting in a nearly continuous spectrum
for $\Delta\gg\hbar\omega_c$. This trend is evident in the results of Fig.~\ref{fig:xxmass}, where we note that the oscillations due to
Landau level spacing are almost completely washed out for $\Delta=1$~eV. Also shown in the figure is the optical conductivity in the 
absence of a magnetic field.\cite{Pedersen2009} Comparing these to the results with magnetic field, it is clear that for a sufficiently large mass term,
$\Delta\gg\hbar\omega_c$, the results are nearly identical, indicating that a large mass term completely cancels the effect of the magnetic field
on the diagonal optical conductivity. We stress that while the cyclotron energy in this case is so large that a mass term of the order 
of $1$~eV is needed to counter the effect of the magnetic field, more realistic values of the magnetic field strength would of course 
lower this threshold significantly.

In general, we find that even at the very large magnetic field of $78$~T used for these calculations, and at chemical potentials
well above the value of the mass term, 
the results for the diagonal conductivity are in excellent agreement between the three methods, so long as
photon energies are well below the $M$-point resonance. We see no reason why such an agreement should break down at lower field
strengths, provided that broadening does not smear out the resonances, 
and we thus argue that the results we obtain will remain qualitatively the same also at more realistic magnetic fields,
with resonances simply red-shifted in accordance with the lower cyclotron energy. 

\subsection{Hall conductivity}
We now turn to the off-diagonal Hall conductivity. Before discussing the optical Hall conductivity, we focus on the DC Hall conductivity
$\sigma_{xy}^0\equiv\sigma_{xy}(0)$ as function of the chemical potential. An intriguing consequence of the linear dispersion relation of graphene is the emergence
of an unconventional quantum Hall effect, with the DC Hall conductivity quantized according to $\sigma_{xy}^0=-(4e^2/h)(n+1/2)$, with
$n$ a positive integer or zero. This unique feature has its origin in the $n=0$ Landau level, the degeneracy of which
is half that of the higher Landau levels.\cite{Gusynin2005,Gusynin2006a} This carries over to the case of gapped graphene, if 
contributions from both valleys are taken into account.\cite{Jiang2010}

\begin{figure}
\begin{center}
\includegraphics[width=\linewidth]{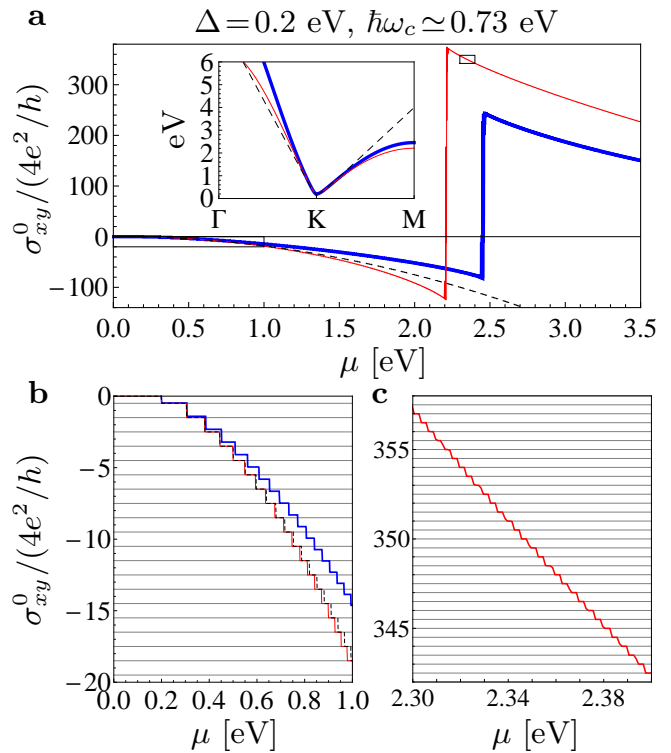}
\caption{(Color online) 
DC Hall conductivity in gapped graphene as a function of the chemical potential. The thick, blue lines are results of the TB model 
including overlap, while thin, red lines are without overlap. Dashed, black lines are the DE results. Note that in contrast to the other figures, 
here the conductivity is given in units of $4e^2/h$. 
(a) DC Hall conductivity in the full range of chemical potentials considered, illustrating the discrepancies between the three models and
the abrupt change in behavior near $\mu=(1+s)|t|$ for the TB model results. 
The inset shows the band structure of gapped graphene without magnetic field.
(b) and (c) offer closer views of the regimes indicated by
black rectangles in panel a. Horizontal lines in (b) indicate $\sigma_{xy}^0=-(4e^2/h)(n+1/2)$, while those in (c) indicate
$\sigma_{xy}^0=(2e^2/h)n$.
}
\label{fig:DCHall}
\end{center}
\end{figure}
In Fig.~\ref{fig:DCHall} we show the DC Hall conductivity as a function of chemical potential, calculated using the three models.
In these calculations, the temperature has been set to $1K$ and the broadening term $\hbar\Gamma$ has been omitted, in order to properly
resolve the quantum Hall plateaus. In panel (a) of the figure, we show the full range of chemical potentials considered.
While there is quite good agreement between all three models for low values of the chemical potential, $\mu\lesssim 0.5$~eV, the breakdown of the Dirac model is
apparent at higher chemical potentials, for which the non-linear part of the band structure is probed. This results in discrepancies
in the energies of the Landau levels, which is apparent in Fig.~\ref{fig:DCHall}b. While the TB model without overlap obviously results in a slightly
different Landau level structure, the values of the DC Hall conductivity still fall at the same plateaus as predicted by the DE model.
The inclusion of overlap alters the results quite significantly, with any but the lowest Hall plateaus falling at non-equidistant values.
We stress that these results are of course calculated for a substantial magnetic field, and we expect much better agreement between the three
models when the Landau level spacing is smaller. The TB results show an abrupt change in behavior near a chemical potential of
$\mu=(1+s)|t|$, where $\sigma_{xy}^0$ suddenly changes sign. This value of the chemical potential coincides with the maximum of the 
band at the $M$ point of gapped graphene without a magnetic field, as illustrated in the inset of Fig.~\ref{fig:DCHall}.
The degeneracy of the Landau levels is abruptly altered at this point, with the four-fold degeneracy for Landau levels below 
$E=(1+s)|t|$ changed to a two-fold degeneracy above this energy, as we have confirmed by inspection of the eigenvalues of both
TB models. This results in a new structure of the quantized Hall plateaus, as shown
in Fig.~\ref{fig:DCHall}c, with the reduced degeneracy manifesting as Hall plateaus at values $\sigma_{xy}^0=(2e^2/h)n$.
As they have their origin in the non-linear part of the band structure, these features are of course entirely absent in the DE results.

We now turn to the optical Hall conductivity. As described previously and illustrated in Fig.~\ref{fig:DEtrans}, 
the DE model predicts an optical Hall conductivity identically zero at zero temperature and for a chemical potential fixed at midgap, 
due to the complete cancellation of conjugated transitions  in opposite valleys. Below, we will show that this does not hold
for the TB model with overlap, but for now we focus on the case where the chemical potential is located at the lowest Landau level, $\mu=\Delta$.
In this case, the symmetry argument obviously breaks down due to the different Fermi distribution functions of the conjugated transitions,
and all three models predict non-zero Hall conductivities.

\begin{figure}
\begin{center}
\includegraphics[width=\linewidth]{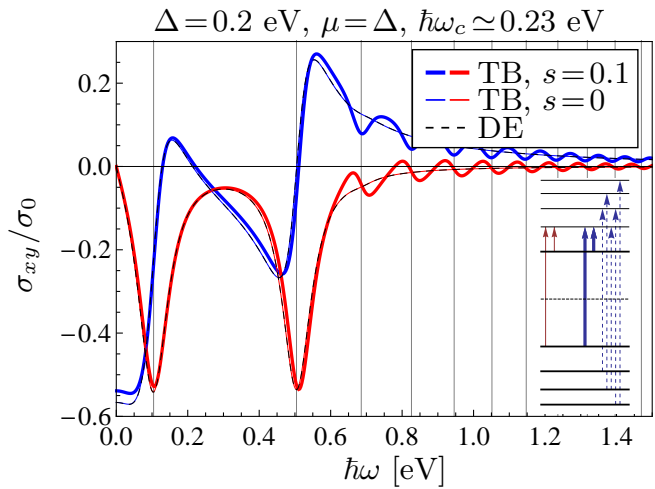}
\caption{(Color online) Optical Hall conductivity $\sigma_{xy}$ in units of the DC graphene
conductivity $\sigma_0$, as a function of photon energy $\hbar\omega$. Results are shown for a magnetic field $B=78$~T. 
The chemical potential is set at the value of the mass term, $\mu=\Delta$.
Thick lines
indicate results from the TB model including overlap, while thin lines are without overlap.
Dashed lines are DE results. Blue (red) coloring indicates the real (imaginary) part of the conductivity.
Vertical lines indicate the resonance energies predicted by the DE model.
The inset illustrates the transitions contributing to the Hall conductivity. See text for more details.
}
\label{fig:xyd0p2mu0p2}
\end{center}
\end{figure}
In Fig.~\ref{fig:xyd0p2mu0p2} we show the optical Hall conductivity of gapped graphene with a mass term of $\Delta=0.2$~eV and the chemical
potential at $\mu=\Delta$. Common to the results of all three models are the two pronounced resonances, that the DE model predict at
$\hbar\omega_{\pm}=\Delta\left(\sqrt{1+\hbar^2\omega_c^2/\Delta^2}\pm 1\right)$, similar to the lowest resonances of the corresponding diagonal
conductivity shown in Fig.~\ref{fig:xxd0p2mu0p2}. However, contrary to the diagonal conductivity, higher-energy resonances are strongly
suppressed in the case of the TB model with overlap, and are entirely absent in the two other models. This can be explained quite straightforwardly
by considering the conjugate transitions and the effect of the overlap on the band structure. We illustrate this in the inset of 
Fig.~\ref{fig:xyd0p2mu0p2}. Here, the thin, red arrows indicate the transitions contributing to the optical Hall conductivity in the case of the DE
model and the TB model without overlap. Because of perfect electron-hole symmetry in both these models (when including both valleys), contributions 
from all other transitions are canceled by their conjugates, the only transitions contributing being those involving the state at $E=\Delta$.
When overlap is included (indicated by thick, blue arrows in the inset), electron-hole symmetry is broken and the conjugate transitions no longer cancel due 
to a slight difference in their transition energies (dashed arrows in the inset). This results in a much richer structure of the optical 
Hall conductivity when including overlap, with additional resonances occurring due to transitions between higher-lying Landau levels. However,
because the energy difference between electron and hole states induced by the overlap is only of the order $s^2$, the strengths of these resonances
are reduced significantly compared to those stemming from the $E=\Delta$ Landau level.

\begin{figure}
\begin{center}
\includegraphics[width=\linewidth]{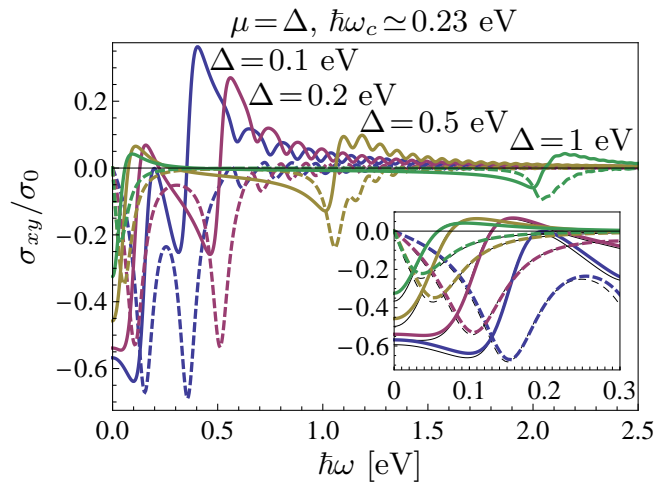}
\caption{(Color online) Optical Hall conductivity $\sigma_{xy}$ in units of the DC graphene
conductivity $\sigma_0$, as a function of photon energy $\hbar\omega$. Results are shown for a magnetic field $B=78$~T
and for four different values of the mass term, all calculated using the TB model with overlap. The chemical potential
is in each case fixed at the lowest Landau level, $\mu=\Delta$. Full (dashed) lines
indicate the real (imaginary) part of the conductivity. 
The inset shows a closer view of the first resonance. The results of the TB model without overlap is shown by
thin, black lines, illustrating the slight discrepancies between the models in this regime. The DE results are practically
identical to those of the TB model without overlap.
}
\label{fig:xyMassTerm}
\end{center}
\end{figure}
In Fig.~\ref{fig:xyMassTerm} we illustrate the effect of the mass term on the optical Hall conductivity. We show results of the TB model with
overlap, and increase the mass term from $\Delta=0.1$~eV to $\Delta=1$~eV, while in each case fixing the chemical potential at the energy of the lowest
Landau level, $\mu=\Delta$. As for the diagonal conductivity, we clearly see that the effect of the magnetic field is strongly reduced when
the mass term dominates the cyclotron energy. Thus, whereas at large values of the mass term the diagonal conductivity retains the resonance 
corresponding to a transition energy of roughly the size of the band gap, this resonance vanishes in the optical Hall conductivity, once the
mass term grows sufficiently large. In the inset of Fig.~\ref{fig:xyMassTerm}, we show a closer view of the first resonance at $E_1-\Delta$,
and include results of the TB model without overlap. While there are some small discrepancies between the two models in this regime,
the overall agreement between the models is quite good. The corresponding results of the DE model are nearly identical to those of the TB model without 
overlap. We note that the DC Hall conductivity is also decreased as the mass term increases, due to squeezing of the Landau level spacing.

As noted above, an interesting consequence of including overlap in the TB model is that the electron-hole symmetry is broken, and
exact cancellation of conjugate transitions at zero chemical potential no longer applies. We therefore expect that the inclusion of overlap
in the TB model will result in a non-zero optical Hall conductivity even at vanishing temperatures and the chemical potential fixed midgap.
\begin{figure}
\begin{center}
\includegraphics[width=\linewidth]{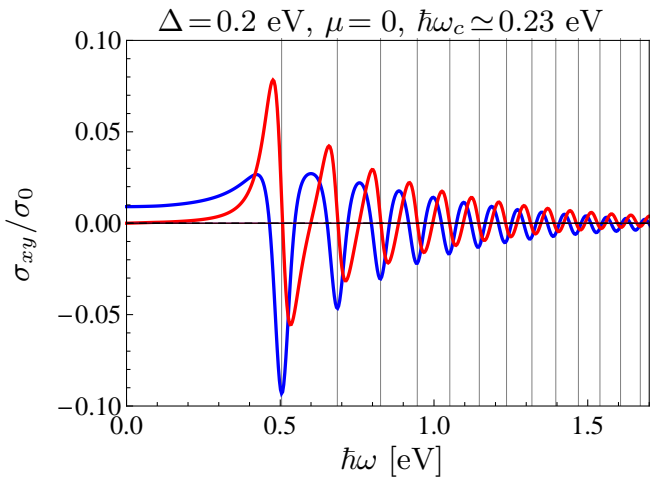}
\caption{(Color online) Optical Hall conductivity with a chemical potential fixed at midgap, calculated using the TB model with overlap.
The vertical lines indicate the transition energies predicted by the DE model.
}
\label{fig:xyd0p2mu0}
\end{center}
\end{figure}
In Fig.~\ref{fig:xyd0p2mu0} we illustrate this, showing the optical Hall conductivity calculated using the TB model with overlap, with a mass term 
$\Delta=0.2$~eV and the chemical potential at midgap. As expected, while the other two methods predict a zero Hall conductivity, the inclusion
of overlap breaks the electron-hole symmetry and results in a non-zero optical Hall conductivity. Overlap is thus a crucial ingredient for properly
evaluating the optical Hall conductivity of graphene. However, we note that because it stems from the overlap, 
the magnitude of the Hall conductivity remains significantly lower than the case where the chemical potential sits at the lowest Landau level.

\section{Summary}\label{Conclusions}
We have calculated the optical conductivity tensor of gapped graphene in presence of a magnetic field, by employing a Peierls substitution in
a nearest-neighbor tight-binding model, both with and without overlap. By generalizing results from ordinary graphene with no magnetic field,
we have found a simple relation between energy eigenvalues calculated with and without overlap, allowing us to avoid solving a large generalized 
eigenvalue problem to account for the overlap. This simplification is valid so long as $s\Delta \ll t$. We have compared the optical conductivities
calculated using the tight-binding model with analytical expressions obtained in a Dirac equation approach.\cite{Gusynin2007} 
To elucidate the differences
between these two approaches we have highlighted the role played by the mass term in redefining the symmetry between electron and hole states in the Dirac
model. This results in a non-zero optical Hall conductivity in individual valleys even with a chemical potential fixed in the middle of the mass gap.
However, summing contributions from both valleys, the symmetry induced by the mass term between electron and hole states in opposite valleys results
in an optical Hall conductivity identically zero. This result carries over to the tight-binding model without overlap, and in general we find
excellent agreement between the optical Hall conductivities calculated with these two methods.


Including overlap in the tight-binding model strongly modifies this picture, because the energy spectrum no
longer has perfect electron-hole symmetry. We find that this results in a much richer structure of the optical Hall conductivity, while also predicting
non-zero optical Hall conductivity even with a chemical potential fixed midgap. We conclude that overlap is a crucial ingredient for a proper, full
description of the off-diagonal magneto-optical properties of gapped graphene.
While the optical Hall conductivity does show some significant discrepancies between the three models, in general we find that the low-energy
transitions are, as expected, very well captured in the simple Dirac model. In particular, the diagonal optical conductivity at transition energies
well below the saddle point resonance at $4.4$~eV is almost identical between all three models.

Finally, we find that a sufficiently large mass term, much larger than the cyclotron energy, effectively washes out the influence of the magnetic field on
the optical properties of gapped graphene.

\begin{acknowledgments}
Financial support from Danish Research Council FTP grant ``Nanoengineered graphene devices'' is gratefully acknowledged.
\end{acknowledgments}


\begin{thebibliography}{35}
\expandafter\ifx\csname natexlab\endcsname\relax\def\natexlab#1{#1}\fi
\expandafter\ifx\csname bibnamefont\endcsname\relax
  \def\bibnamefont#1{#1}\fi
\expandafter\ifx\csname bibfnamefont\endcsname\relax
  \def\bibfnamefont#1{#1}\fi
\expandafter\ifx\csname citenamefont\endcsname\relax
  \def\citenamefont#1{#1}\fi
\expandafter\ifx\csname url\endcsname\relax
  \def\url#1{\texttt{#1}}\fi
\expandafter\ifx\csname urlprefix\endcsname\relax\def\urlprefix{URL }\fi
\providecommand{\bibinfo}[2]{#2}
\providecommand{\eprint}[2][]{\url{#2}}

\bibitem[{\citenamefont{Geim and Novoselov}(2007)}]{Geim2007}
\bibinfo{author}{\bibfnamefont{A.~K.} \bibnamefont{Geim}} \bibnamefont{and}
  \bibinfo{author}{\bibfnamefont{K.~S.} \bibnamefont{Novoselov}},
  \bibinfo{journal}{Nature Materials} \textbf{\bibinfo{volume}{6}},
  \bibinfo{pages}{183} (\bibinfo{year}{2007}).

\bibitem[{\citenamefont{Geim}(2009)}]{Geim2009}
\bibinfo{author}{\bibfnamefont{A.~K.} \bibnamefont{Geim}},
  \bibinfo{journal}{Science} \textbf{\bibinfo{volume}{19}},
  \bibinfo{pages}{1530} (\bibinfo{year}{2009}).

\bibitem[{\citenamefont{Novoselov et~al.}(2004)\citenamefont{Novoselov, Geim,
  Morozov, Jiang, Zhang, Dubonos, Grigorieva, and Firsov}}]{Novoselov2004}
\bibinfo{author}{\bibfnamefont{K.~S.} \bibnamefont{Novoselov}},
  \bibinfo{author}{\bibfnamefont{A.~K.} \bibnamefont{Geim}},
  \bibinfo{author}{\bibfnamefont{S.~V.} \bibnamefont{Morozov}},
  \bibinfo{author}{\bibfnamefont{D.}~\bibnamefont{Jiang}},
  \bibinfo{author}{\bibfnamefont{Y.}~\bibnamefont{Zhang}},
  \bibinfo{author}{\bibfnamefont{S.~V.} \bibnamefont{Dubonos}},
  \bibinfo{author}{\bibfnamefont{I.~V.} \bibnamefont{Grigorieva}},
  \bibnamefont{and} \bibinfo{author}{\bibfnamefont{A.~A.}
  \bibnamefont{Firsov}}, \bibinfo{journal}{Science}
  \textbf{\bibinfo{volume}{306}}, \bibinfo{pages}{666} (\bibinfo{year}{2004}).

\bibitem[{\citenamefont{{Castro Neto} et~al.}(2009)\citenamefont{{Castro Neto},
  Guinea, Peres, Novoselov, and Geim}}]{Neto2009}
\bibinfo{author}{\bibfnamefont{A.~H.} \bibnamefont{{Castro Neto}}},
  \bibinfo{author}{\bibfnamefont{F.}~\bibnamefont{Guinea}},
  \bibinfo{author}{\bibfnamefont{N.~M.~R.} \bibnamefont{Peres}},
  \bibinfo{author}{\bibfnamefont{K.~S.} \bibnamefont{Novoselov}},
  \bibnamefont{and} \bibinfo{author}{\bibfnamefont{A.~K.} \bibnamefont{Geim}},
  \bibinfo{journal}{Rev. Mod. Phys.} \textbf{\bibinfo{volume}{81}},
  \bibinfo{pages}{109} (\bibinfo{year}{2009}).

\bibitem[{\citenamefont{Morozov et~al.}(2008)\citenamefont{Morozov, Novoselov,
  Katsnelson, Schedin, Elias, Jaszczak, and Geim}}]{Morozov2008}
\bibinfo{author}{\bibfnamefont{S.~V.} \bibnamefont{Morozov}},
  \bibinfo{author}{\bibfnamefont{K.~S.} \bibnamefont{Novoselov}},
  \bibinfo{author}{\bibfnamefont{M.~I.} \bibnamefont{Katsnelson}},
  \bibinfo{author}{\bibfnamefont{F.}~\bibnamefont{Schedin}},
  \bibinfo{author}{\bibfnamefont{D.~C.} \bibnamefont{Elias}},
  \bibinfo{author}{\bibfnamefont{J.~A.} \bibnamefont{Jaszczak}},
  \bibnamefont{and} \bibinfo{author}{\bibfnamefont{A.~K.} \bibnamefont{Geim}},
  \bibinfo{journal}{Phys. Rev. Lett.} \textbf{\bibinfo{volume}{100}},
  \bibinfo{pages}{016602} (\bibinfo{year}{2008}).

\bibitem[{\citenamefont{Novoselov et~al.}(2007)\citenamefont{Novoselov, Jiang,
  Zhang, Morozov, Stormer, Zeitler, Maan, Boebinger, Kim, and
  Geim}}]{Novoselov2007}
\bibinfo{author}{\bibfnamefont{K.~S.} \bibnamefont{Novoselov}},
  \bibinfo{author}{\bibfnamefont{Z.}~\bibnamefont{Jiang}},
  \bibinfo{author}{\bibfnamefont{Y.}~\bibnamefont{Zhang}},
  \bibinfo{author}{\bibfnamefont{S.~V.} \bibnamefont{Morozov}},
  \bibinfo{author}{\bibfnamefont{H.~L.} \bibnamefont{Stormer}},
  \bibinfo{author}{\bibfnamefont{U.}~\bibnamefont{Zeitler}},
  \bibinfo{author}{\bibfnamefont{J.~C.} \bibnamefont{Maan}},
  \bibinfo{author}{\bibfnamefont{G.~S.} \bibnamefont{Boebinger}},
  \bibinfo{author}{\bibfnamefont{P.}~\bibnamefont{Kim}}, \bibnamefont{and}
  \bibinfo{author}{\bibfnamefont{A.~K.} \bibnamefont{Geim}},
  \bibinfo{journal}{Science} \textbf{\bibinfo{volume}{315}},
  \bibinfo{pages}{1379} (\bibinfo{year}{2007}).

\bibitem[{\citenamefont{Novoselov et~al.}(2005)\citenamefont{Novoselov, Geim,
  Morozov, Jiang, Katsnelson, Grigorieva, Dubonos, and Firsov}}]{Novoselov2005}
\bibinfo{author}{\bibfnamefont{K.~S.} \bibnamefont{Novoselov}},
  \bibinfo{author}{\bibfnamefont{A.~K.} \bibnamefont{Geim}},
  \bibinfo{author}{\bibfnamefont{S.~V.} \bibnamefont{Morozov}},
  \bibinfo{author}{\bibfnamefont{D.}~\bibnamefont{Jiang}},
  \bibinfo{author}{\bibfnamefont{M.~I.} \bibnamefont{Katsnelson}},
  \bibinfo{author}{\bibfnamefont{I.~V.} \bibnamefont{Grigorieva}},
  \bibinfo{author}{\bibfnamefont{S.~V.} \bibnamefont{Dubonos}},
  \bibnamefont{and} \bibinfo{author}{\bibfnamefont{A.~A.}
  \bibnamefont{Firsov}}, \bibinfo{journal}{Nature}
  \textbf{\bibinfo{volume}{438}}, \bibinfo{pages}{197} (\bibinfo{year}{2005}).

\bibitem[{\citenamefont{Zhang et~al.}(2005)\citenamefont{Zhang, Tan, Stormer,
  and Kim}}]{Zhang2005}
\bibinfo{author}{\bibfnamefont{Y.}~\bibnamefont{Zhang}},
  \bibinfo{author}{\bibfnamefont{Y.-W.} \bibnamefont{Tan}},
  \bibinfo{author}{\bibfnamefont{H.~L.} \bibnamefont{Stormer}},
  \bibnamefont{and} \bibinfo{author}{\bibfnamefont{P.}~\bibnamefont{Kim}},
  \bibinfo{journal}{Nature} \textbf{\bibinfo{volume}{438}},
  \bibinfo{pages}{201} (\bibinfo{year}{2005}).

\bibitem[{\citenamefont{Gusynin and
  Sharapov}(2005{\natexlab{a}})}]{Gusynin2005}
\bibinfo{author}{\bibfnamefont{V.~P.} \bibnamefont{Gusynin}} \bibnamefont{and}
  \bibinfo{author}{\bibfnamefont{S.~G.} \bibnamefont{Sharapov}},
  \bibinfo{journal}{Phys. Rev. Lett.} \textbf{\bibinfo{volume}{95}},
  \bibinfo{pages}{146801} (\bibinfo{year}{2005}{\natexlab{a}}).

\bibitem[{\citenamefont{Gusynin and
  Sharapov}(2005{\natexlab{b}})}]{Gusynin2005a}
\bibinfo{author}{\bibfnamefont{V.~P.} \bibnamefont{Gusynin}} \bibnamefont{and}
  \bibinfo{author}{\bibfnamefont{S.~G.} \bibnamefont{Sharapov}},
  \bibinfo{journal}{Phys. Rev. B} \textbf{\bibinfo{volume}{71}},
  \bibinfo{pages}{125124} (\bibinfo{year}{2005}{\natexlab{b}}).

\bibitem[{\citenamefont{Lin et~al.}(2010)\citenamefont{Lin, Dimitrakopoulos,
  Jenkins, Farmer, Chiu, Grill, and Avouris}}]{Lin2010}
\bibinfo{author}{\bibfnamefont{Y.-M.} \bibnamefont{Lin}},
  \bibinfo{author}{\bibfnamefont{C.}~\bibnamefont{Dimitrakopoulos}},
  \bibinfo{author}{\bibfnamefont{K.~A.} \bibnamefont{Jenkins}},
  \bibinfo{author}{\bibfnamefont{D.~B.} \bibnamefont{Farmer}},
  \bibinfo{author}{\bibfnamefont{H.-Y.} \bibnamefont{Chiu}},
  \bibinfo{author}{\bibfnamefont{A.}~\bibnamefont{Grill}}, \bibnamefont{and}
  \bibinfo{author}{\bibfnamefont{P.}~\bibnamefont{Avouris}},
  \bibinfo{journal}{Science} \textbf{\bibinfo{volume}{327}},
  \bibinfo{pages}{662} (\bibinfo{year}{2010}).

\bibitem[{\citenamefont{Park et~al.}(2010)\citenamefont{Park, Kim, Bulovic, and
  Kong}}]{Park2010}
\bibinfo{author}{\bibfnamefont{H.}~\bibnamefont{Park}},
  \bibinfo{author}{\bibfnamefont{J.~A. R.~K.} \bibnamefont{Kim}},
  \bibinfo{author}{\bibfnamefont{V.}~\bibnamefont{Bulovic}}, \bibnamefont{and}
  \bibinfo{author}{\bibfnamefont{J.}~\bibnamefont{Kong}},
  \bibinfo{journal}{Nanotechnology} \textbf{\bibinfo{volume}{21}},
  \bibinfo{pages}{505204} (\bibinfo{year}{2010}).

\bibitem[{\citenamefont{Patchkovskii et~al.}(2005)\citenamefont{Patchkovskii,
  Tse, Yurchenko, Zhechkov, Heine, and Seifert}}]{Patchkovskii2005}
\bibinfo{author}{\bibfnamefont{S.}~\bibnamefont{Patchkovskii}},
  \bibinfo{author}{\bibfnamefont{J.~S.} \bibnamefont{Tse}},
  \bibinfo{author}{\bibfnamefont{S.~N.} \bibnamefont{Yurchenko}},
  \bibinfo{author}{\bibfnamefont{L.}~\bibnamefont{Zhechkov}},
  \bibinfo{author}{\bibfnamefont{T.}~\bibnamefont{Heine}}, \bibnamefont{and}
  \bibinfo{author}{\bibfnamefont{G.}~\bibnamefont{Seifert}},
  \bibinfo{journal}{Proc. Natl. Acad. Sci. USA} \textbf{\bibinfo{volume}{102}},
  \bibinfo{pages}{10439} (\bibinfo{year}{2005}).

\bibitem[{\citenamefont{Bae et~al.}(2010)\citenamefont{Bae, Kim, Lee, Xu, Park,
  Zheng, Balakrishnan, Lei, Ri~Kim, Song et~al.}}]{Bae2010}
\bibinfo{author}{\bibfnamefont{S.}~\bibnamefont{Bae}},
  \bibinfo{author}{\bibfnamefont{H.}~\bibnamefont{Kim}},
  \bibinfo{author}{\bibfnamefont{Y.}~\bibnamefont{Lee}},
  \bibinfo{author}{\bibfnamefont{X.}~\bibnamefont{Xu}},
  \bibinfo{author}{\bibfnamefont{J.-S.} \bibnamefont{Park}},
  \bibinfo{author}{\bibfnamefont{Y.}~\bibnamefont{Zheng}},
  \bibinfo{author}{\bibfnamefont{J.}~\bibnamefont{Balakrishnan}},
  \bibinfo{author}{\bibfnamefont{T.}~\bibnamefont{Lei}},
  \bibinfo{author}{\bibfnamefont{H.}~\bibnamefont{Ri~Kim}},
  \bibinfo{author}{\bibfnamefont{Y.~I.} \bibnamefont{Song}},
  \bibnamefont{et~al.}, \bibinfo{journal}{Nat. Nano.}
  \textbf{\bibinfo{volume}{5}}, \bibinfo{pages}{574} (\bibinfo{year}{2010}).

\bibitem[{\citenamefont{Nakada et~al.}(1996)\citenamefont{Nakada, Fujita,
  Dresselhaus, and Dresselhaus}}]{Nakada1996}
\bibinfo{author}{\bibfnamefont{K.}~\bibnamefont{Nakada}},
  \bibinfo{author}{\bibfnamefont{M.}~\bibnamefont{Fujita}},
  \bibinfo{author}{\bibfnamefont{G.}~\bibnamefont{Dresselhaus}},
  \bibnamefont{and} \bibinfo{author}{\bibfnamefont{M.~S.}
  \bibnamefont{Dresselhaus}}, \bibinfo{journal}{Phys. Rev. B}
  \textbf{\bibinfo{volume}{54}}, \bibinfo{pages}{17954} (\bibinfo{year}{1996}).

\bibitem[{\citenamefont{Son et~al.}(2006)\citenamefont{Son, Cohen, and
  Louie}}]{Son2006}
\bibinfo{author}{\bibfnamefont{Y.-W.} \bibnamefont{Son}},
  \bibinfo{author}{\bibfnamefont{M.~L.} \bibnamefont{Cohen}}, \bibnamefont{and}
  \bibinfo{author}{\bibfnamefont{S.~G.} \bibnamefont{Louie}},
  \bibinfo{journal}{Phys. Rev. Lett.} \textbf{\bibinfo{volume}{97}},
  \bibinfo{pages}{216803} (\bibinfo{year}{2006}).

\bibitem[{\citenamefont{Brey and Fertig}(2006)}]{Brey2006}
\bibinfo{author}{\bibfnamefont{L.}~\bibnamefont{Brey}} \bibnamefont{and}
  \bibinfo{author}{\bibfnamefont{H.~A.} \bibnamefont{Fertig}},
  \bibinfo{journal}{Phys. Rev. B} \textbf{\bibinfo{volume}{73}},
  \bibinfo{pages}{235411} (\bibinfo{year}{2006}).

\bibitem[{\citenamefont{Sofo et~al.}(2007)\citenamefont{Sofo, Chaudhari, and
  Barber}}]{Sofo2007}
\bibinfo{author}{\bibfnamefont{J.~O.} \bibnamefont{Sofo}},
  \bibinfo{author}{\bibfnamefont{A.~S.} \bibnamefont{Chaudhari}},
  \bibnamefont{and} \bibinfo{author}{\bibfnamefont{G.~D.}
  \bibnamefont{Barber}}, \bibinfo{journal}{Physical Review B}
  \textbf{\bibinfo{volume}{75}}, \bibinfo{pages}{153401}
  (\bibinfo{year}{2007}).

\bibitem[{\citenamefont{Elias et~al.}(2009)\citenamefont{Elias, Nair,
  Mohiuddin, Morozov, Blake, Halsall, Ferrari, Boukhvalov, Katsnelson, Geim
  et~al.}}]{Elias2009}
\bibinfo{author}{\bibfnamefont{D.~C.} \bibnamefont{Elias}},
  \bibinfo{author}{\bibfnamefont{R.~R.} \bibnamefont{Nair}},
  \bibinfo{author}{\bibfnamefont{T.~M.~G.} \bibnamefont{Mohiuddin}},
  \bibinfo{author}{\bibfnamefont{S.~V.} \bibnamefont{Morozov}},
  \bibinfo{author}{\bibfnamefont{P.}~\bibnamefont{Blake}},
  \bibinfo{author}{\bibfnamefont{M.~P.} \bibnamefont{Halsall}},
  \bibinfo{author}{\bibfnamefont{A.~C.} \bibnamefont{Ferrari}},
  \bibinfo{author}{\bibfnamefont{D.~W.} \bibnamefont{Boukhvalov}},
  \bibinfo{author}{\bibfnamefont{M.~I.} \bibnamefont{Katsnelson}},
  \bibinfo{author}{\bibfnamefont{A.~K.} \bibnamefont{Geim}},
  \bibnamefont{et~al.}, \bibinfo{journal}{Science}
  \textbf{\bibinfo{volume}{323}}, \bibinfo{pages}{610} (\bibinfo{year}{2009}).

\bibitem[{\citenamefont{Pedersen
  et~al.}(2008{\natexlab{a}})\citenamefont{Pedersen, Flindt, Pedersen,
  Mortensen, Jauho, and Pedersen}}]{Pedersen2008}
\bibinfo{author}{\bibfnamefont{T.~G.} \bibnamefont{Pedersen}},
  \bibinfo{author}{\bibfnamefont{C.}~\bibnamefont{Flindt}},
  \bibinfo{author}{\bibfnamefont{J.}~\bibnamefont{Pedersen}},
  \bibinfo{author}{\bibfnamefont{N.~A.} \bibnamefont{Mortensen}},
  \bibinfo{author}{\bibfnamefont{A.-P.} \bibnamefont{Jauho}}, \bibnamefont{and}
  \bibinfo{author}{\bibfnamefont{K.}~\bibnamefont{Pedersen}},
  \bibinfo{journal}{Phys. Rev. Lett.} \textbf{\bibinfo{volume}{100}},
  \bibinfo{pages}{136804} (\bibinfo{year}{2008}{\natexlab{a}}).

\bibitem[{\citenamefont{Pedersen
  et~al.}(2008{\natexlab{b}})\citenamefont{Pedersen, Flindt, Pedersen, Jauho,
  Mortensen, and Pedersen}}]{Pedersen2008a}
\bibinfo{author}{\bibfnamefont{T.~G.} \bibnamefont{Pedersen}},
  \bibinfo{author}{\bibfnamefont{C.}~\bibnamefont{Flindt}},
  \bibinfo{author}{\bibfnamefont{J.}~\bibnamefont{Pedersen}},
  \bibinfo{author}{\bibfnamefont{A.-P.} \bibnamefont{Jauho}},
  \bibinfo{author}{\bibfnamefont{N.~A.} \bibnamefont{Mortensen}},
  \bibnamefont{and} \bibinfo{author}{\bibfnamefont{K.}~\bibnamefont{Pedersen}},
  \bibinfo{journal}{Phys. Rev. B} \textbf{\bibinfo{volume}{77}},
  \bibinfo{pages}{245431} (\bibinfo{year}{2008}{\natexlab{b}}).

\bibitem[{\citenamefont{Balog et~al.}(2010)\citenamefont{Balog, Jørgensen,
  Nilsson, Andersen, Rienks, Bianchi, Fanetti, Lægsgaard, Baraldi, Lizzit
  et~al.}}]{Balog2010}
\bibinfo{author}{\bibfnamefont{R.}~\bibnamefont{Balog}},
  \bibinfo{author}{\bibfnamefont{B.}~\bibnamefont{Jørgensen}},
  \bibinfo{author}{\bibfnamefont{L.}~\bibnamefont{Nilsson}},
  \bibinfo{author}{\bibfnamefont{M.}~\bibnamefont{Andersen}},
  \bibinfo{author}{\bibfnamefont{E.}~\bibnamefont{Rienks}},
  \bibinfo{author}{\bibfnamefont{M.}~\bibnamefont{Bianchi}},
  \bibinfo{author}{\bibfnamefont{M.}~\bibnamefont{Fanetti}},
  \bibinfo{author}{\bibfnamefont{E.}~\bibnamefont{Lægsgaard}},
  \bibinfo{author}{\bibfnamefont{A.}~\bibnamefont{Baraldi}},
  \bibinfo{author}{\bibfnamefont{S.}~\bibnamefont{Lizzit}},
  \bibnamefont{et~al.}, \bibinfo{journal}{Nature Materials}
  \textbf{\bibinfo{volume}{9}}, \bibinfo{pages}{315} (\bibinfo{year}{2010}).

\bibitem[{\citenamefont{Bai et~al.}(2010)\citenamefont{Bai, Zhong, Jiang,
  Huang, and Duan}}]{Bai2010}
\bibinfo{author}{\bibfnamefont{J.}~\bibnamefont{Bai}},
  \bibinfo{author}{\bibfnamefont{X.}~\bibnamefont{Zhong}},
  \bibinfo{author}{\bibfnamefont{S.}~\bibnamefont{Jiang}},
  \bibinfo{author}{\bibfnamefont{Y.}~\bibnamefont{Huang}}, \bibnamefont{and}
  \bibinfo{author}{\bibfnamefont{X.}~\bibnamefont{Duan}},
  \bibinfo{journal}{Nature Nanotechnology} \textbf{\bibinfo{volume}{5}},
  \bibinfo{pages}{190} (\bibinfo{year}{2010}).

\bibitem[{\citenamefont{Kim et~al.}(2010)\citenamefont{Kim, Safron, Han,
  Arnold, and Gopalan}}]{Kim2010}
\bibinfo{author}{\bibfnamefont{M.}~\bibnamefont{Kim}},
  \bibinfo{author}{\bibfnamefont{N.~S.} \bibnamefont{Safron}},
  \bibinfo{author}{\bibfnamefont{E.}~\bibnamefont{Han}},
  \bibinfo{author}{\bibfnamefont{M.~S.} \bibnamefont{Arnold}},
  \bibnamefont{and} \bibinfo{author}{\bibfnamefont{P.}~\bibnamefont{Gopalan}},
  \bibinfo{journal}{Nano Lett.} \textbf{\bibinfo{volume}{10}},
  \bibinfo{pages}{1125} (\bibinfo{year}{2010}).

\bibitem[{\citenamefont{Semenoff}(1984)}]{Semenoff1984}
\bibinfo{author}{\bibfnamefont{G.~W.} \bibnamefont{Semenoff}},
  \bibinfo{journal}{Phys. Rev. Lett.} \textbf{\bibinfo{volume}{53}},
  \bibinfo{pages}{2449} (\bibinfo{year}{1984}).

\bibitem[{\citenamefont{Gusynin et~al.}(2006)\citenamefont{Gusynin, Sharapov,
  and Carbotte}}]{Gusynin2006}
\bibinfo{author}{\bibfnamefont{V.~P.} \bibnamefont{Gusynin}},
  \bibinfo{author}{\bibfnamefont{S.~G.} \bibnamefont{Sharapov}},
  \bibnamefont{and} \bibinfo{author}{\bibfnamefont{J.~P.}
  \bibnamefont{Carbotte}}, \bibinfo{journal}{Phys. Rev. Lett.}
  \textbf{\bibinfo{volume}{96}}, \bibinfo{pages}{256802}
  (\bibinfo{year}{2006}).

\bibitem[{\citenamefont{Gusynin et~al.}(2007)\citenamefont{Gusynin, Sharapov,
  and Carbotte}}]{Gusynin2007}
\bibinfo{author}{\bibfnamefont{V.~P.} \bibnamefont{Gusynin}},
  \bibinfo{author}{\bibfnamefont{S.~G.} \bibnamefont{Sharapov}},
  \bibnamefont{and} \bibinfo{author}{\bibfnamefont{J.~P.}
  \bibnamefont{Carbotte}}, \bibinfo{journal}{J. Phys. Cond Mat.}
  \textbf{\bibinfo{volume}{19}}, \bibinfo{pages}{026222}
  (\bibinfo{year}{2007}).

\bibitem[{\citenamefont{Jiang et~al.}(2010)\citenamefont{Jiang, Zheng, Li, and
  Shen}}]{Jiang2010}
\bibinfo{author}{\bibfnamefont{L.}~\bibnamefont{Jiang}},
  \bibinfo{author}{\bibfnamefont{Y.}~\bibnamefont{Zheng}},
  \bibinfo{author}{\bibfnamefont{H.}~\bibnamefont{Li}}, \bibnamefont{and}
  \bibinfo{author}{\bibfnamefont{H.}~\bibnamefont{Shen}},
  \bibinfo{journal}{Nanotechnology} \textbf{\bibinfo{volume}{21}},
  \bibinfo{pages}{145703} (\bibinfo{year}{2010}).

\bibitem[{\citenamefont{Taft}(1964)}]{Taft1964}
\bibinfo{author}{\bibfnamefont{E.~A.} \bibnamefont{Taft}} \bibnamefont{and}
	\bibinfo{author}{\bibfnamefont{H.~R.} \bibnamefont{Phillip}},
  \bibinfo{journal}{Phys. Rev.} \textbf{\bibinfo{volume}{138}},
  \bibinfo{pages}{A197} (\bibinfo{year}{1964}).

\bibitem[{\citenamefont{Pedersen}(2003)}]{Pedersen2003}
\bibinfo{author}{\bibfnamefont{T.~G.} \bibnamefont{Pedersen}},
  \bibinfo{journal}{Phys. Rev. B} \textbf{\bibinfo{volume}{68}},
  \bibinfo{pages}{245104} (\bibinfo{year}{2003}).

\bibitem[{\citenamefont{Luttinger}(1951)}]{Luttinger1951}
\bibinfo{author}{\bibfnamefont{J.~M.} \bibnamefont{Luttinger}},
  \bibinfo{journal}{Phys. Rev.} \textbf{\bibinfo{volume}{84}},
  \bibinfo{pages}{814} (\bibinfo{year}{1951}).

\bibitem[{\citenamefont{Sandu}(2005)}]{Sandu2005}
\bibinfo{author}{\bibfnamefont{T.}~\bibnamefont{Sandu}},
  \bibinfo{journal}{Phys. Rev. B} \textbf{\bibinfo{volume}{72}},
  \bibinfo{pages}{125105} (\bibinfo{year}{2005}).

\bibitem[{\citenamefont{Pedersen
  et~al.}(2009{\natexlab{b}})\citenamefont{Pedersen, Jauho, 
  and Pedersen}}]{Pedersen2009}
\bibinfo{author}{\bibfnamefont{T.~G.} \bibnamefont{Pedersen}},
  \bibinfo{author}{\bibfnamefont{A.-P.} \bibnamefont{Jauho}},
  \bibnamefont{and} \bibinfo{author}{\bibfnamefont{K.}~\bibnamefont{Pedersen}},
  \bibinfo{journal}{Phys. Rev. B} \textbf{\bibinfo{volume}{79}},
  \bibinfo{pages}{113406} (\bibinfo{year}{2009}{\natexlab{b}}).

\bibitem[{\citenamefont{Gusynin and Sharapov}(2006)}]{Gusynin2006a}
\bibinfo{author}{\bibfnamefont{V.~P.} \bibnamefont{Gusynin}} \bibnamefont{and}
  \bibinfo{author}{\bibfnamefont{S.~G.} \bibnamefont{Sharapov}},
  \bibinfo{journal}{Phys. Rev. B} \textbf{\bibinfo{volume}{73}},
  \bibinfo{pages}{245411} (\bibinfo{year}{2006}).

\bibitem[{\citenamefont{Xiao et~al.}(2007)\citenamefont{Xiao, Yao, and
  Niu}}]{Xiao2007}
\bibinfo{author}{\bibfnamefont{D.}~\bibnamefont{Xiao}},
  \bibinfo{author}{\bibfnamefont{W.}~\bibnamefont{Yao}}, \bibnamefont{and}
  \bibinfo{author}{\bibfnamefont{Q.}~\bibnamefont{Niu}},
  \bibinfo{journal}{Phys. Rev. Lett.} \textbf{\bibinfo{volume}{99}},
  \bibinfo{pages}{236809} (\bibinfo{year}{2007}).

\bibitem[{\citenamefont{Rycerz et~al.}(2007)\citenamefont{Rycerz, Tworzyd{\l}o,
  and Beenakker}}]{Rycerz2007}
\bibinfo{author}{\bibfnamefont{A.}~\bibnamefont{Rycerz}},
  \bibinfo{author}{\bibfnamefont{J.}~\bibnamefont{Tworzyd{\l}o}},
  \bibnamefont{and} \bibinfo{author}{\bibfnamefont{C.~J.}
  \bibnamefont{Beenakker}}, \bibinfo{journal}{Nature Physics}
  \textbf{\bibinfo{volume}{3}}, \bibinfo{pages}{172} (\bibinfo{year}{2007}).

\bibitem[{\citenamefont{Gunlycke and White}(2011)}]{Gunlycke2011}
\bibinfo{author}{\bibfnamefont{D.}~\bibnamefont{Gunlycke}} \bibnamefont{and}
  \bibinfo{author}{\bibfnamefont{C.T.}~\bibnamefont{White}},
  \bibinfo{journal}{Phys. Rev. L} \textbf{\bibinfo{volume}{106}},
  \bibinfo{pages}{136806} (\bibinfo{year}{2011}).

\end{thebibliography}
\end{document}